# Evaluating the Energy Efficiency of NPU-Accelerated Machine Learning Inference on Embedded Microcontrollers

Anastasios Fanariotis, Theofanis Orphanoudakis, and Vasilis Fotopoulos

*Abstract*— The deployment of machine learning (ML) models on microcontrollers (MCUs) is constrained by strict energy, latency, and memory requirements, particularly in battery-operated and real-time edge devices. While software-level optimizations such as quantization and pruning reduce model size and computation, hardware acceleration has emerged as a decisive enabler for efficient embedded inference. This paper evaluates the impact of Neural Processing Units (NPUs) on MCU-based ML execution, using the ARM Cortex-M55 core combined with the Ethos-U55 NPU on the Alif Semiconductor Ensemble E7 development board as a representative platform. A rigorous measurement methodology was employed, incorporating per-inference net energy accounting via GPIO-triggered high-resolution digital multimeter synchronization and idle-state subtraction, ensuring accurate attribution of energy costs. Experimental results across six representative ML models—including MiniResNet, MobileNetV2, FD-MobileNet, MNIST, TinyYolo, and SSD-MobileNet—demonstrate substantial efficiency gains when inference is offloaded to the NPU. For moderate to large networks, latency improvements ranged from 7× to over 125×, with per-inference net energy reductions up to 143×. Notably, the NPU enabled execution of models unsupported on CPU-only paths, such as SSD-MobileNet, highlighting its functional as well as efficiency advantages. These findings establish NPUs as a cornerstone of energy-aware embedded AI, enabling real-time, power-constrained ML inference at the MCU level.

## I. INTRODUCTION

Machine learning (ML) inference on embedded microcontrollers (MCUs) has emerged as a critical enabler of pervasive intelligence at the edge, powering applications in wearable devices, autonomous sensing, industrial monitoring, and IoT nodes. However, the deployment of deep learning models in these environments faces formidable challenges. MCUs are inherently resource-constrained, with limited RAM, modest Flash capacity, narrow memory bandwidth, and stringent power envelopes. As a result, direct execution of convolutional and fully connected networks on MCU cores often results in prohibitive inference latency and excessive energy consumption, restricting practical deployment to only the simplest classifiers.

To address these constraints, extensive research has focused on software-level optimization techniques such as quantization, pruning, knowledge distillation, and factorization, which reduce computational load and memory footprint. While effective, these methods alone are insufficient to achieve the levels of efficiency required for real-time, battery-powered operation. Hardware acceleration has thus become indispensable, encompassing general-purpose enhancements such as SIMD vectoring and caching hierarchies, as well as dedicated accelerators such as Neural Processing Units (NPUs). These mechanisms provide varying levels of performance scaling, with NPUs representing the most specialized and energy-efficient solution for executing modern convolutional neural networks (CNNs) in MCU-class devices.

Despite the promise of NPUs, quantitative, per-inference evaluation of their real-world benefits on MCUs remains limited in the literature. Many prior studies report peak throughput or theoretical performance figures, but fail to capture practical efficiency under constrained execution environments. In particular, the metric of per-inference net energy, which combines instantaneous power with inference latency, provides a more accurate and application-relevant measure of efficiency for embedded systems. By accounting for both energy draw and execution time, this metric differentiates between superficially efficient but long-latency models and architectures that achieve true energy proportionality.

This paper addresses this gap by presenting a comprehensive measurement and evaluation of NPU acceleration in microcontrollers. Specifically, we employ the Alif Ensemble E7 development board, featuring an ARM Cortex-M55 CPU paired with the Ethos-U55 NPU, and benchmark multiple neural network models spanning different complexity levels. Our contributions are threefold:
- We introduce a rigorous measurement methodology based on per-inference net energy, incorporating high-precision external instrumentation synchronized via

Anastasios Fanariotis is with the Digital Systems and media Computing Laboratory, School of Science and technology, Hellenic Open University, Patras, Greece (email: afanariotis@eap.gr).

Vasilis Fotopoulos is with the Digital Systems and media Computing Laboratory, School of Science and technology, Hellenic Open University, Patras, Greece (email: vfotop1@eap.gr@eap.gr).

Theofanis Orphanoudakis is with the Digital Systems and media Computing Laboratory, School of Science and technology, Hellenic Open University, Patras, Greece (email: fanis@eap.gr).

GPIO triggers to isolate inference energy consumption from idle-state power.
- We present a detailed quantitative comparison of CPU-only versus NPU-assisted inference across six representative ML models, analyzing latency, per-inference energy, memory footprint, and NPU utilization.
- We provide an evidence-based discussion on the role of NPUs within the broader landscape of embedded hardware acceleration, highlighting their superiority for moderate to large models and their indispensability for executing models otherwise infeasible on MCU cores.

The remainder of this paper is structured as follows. Section II reviews background and related work on embedded ML acceleration. Section III details the measurement methodology employed. Section IV describes NPU architectures and their key design strategies. Section V presents the experimental results, including latency, energy, and utilization benchmarks. Section VI discusses the implications of these findings for embedded system design. Section VII concludes with design guidelines and outlines future research directions.

## II. BACKGROUND AND RELATED WORK

Machine learning inference on embedded systems has traditionally been limited by resource constraints, requiring careful balancing between accuracy, latency, and energy consumption. Early approaches to efficient edge inference relied on software-level optimization techniques, including quantization, pruning, and factorization. Quantization reduces the numerical precision of weights and activations, thereby lowering memory and compute demands [1]. Structured pruning selectively removes redundant connections, reducing model size and execution complexity [2]. Factorization and low-rank approximations further decompose large tensors into smaller, computationally efficient forms [3]. These methods have demonstrated significant improvements in model compactness and execution feasibility on microcontrollers; however, they typically involve trade-offs in accuracy and provide limited relief against memory-bandwidth bottlenecks.

To complement algorithmic optimizations, general-purpose hardware acceleration mechanisms have been incorporated into modern microcontroller architectures. Examples include SIMD (Single Instruction Multiple Data) vector extensions, such as ARM's M-Profile Vector Extension (MVE/Helium), or similar vector instruction sets in platforms like Espressif's LX7 cores. These extensions exploit fine-grained parallelism in neural network operators, particularly matrix multiplications and convolutions, leading to consistent but moderate performance gains. Another widely adopted technique is the inclusion of hierarchical memory structures and caching, which minimize high-latency off-chip memory accesses and improve data locality. Prior studies show that caching and vectoring combined yield multiplicative improvements in energy efficiency for memory-bound workloads [4], [5].

In parallel, the emergence of dedicated neural accelerators has redefined the performance landscape of embedded AI. Neural Processing Units (NPUs) integrate specialized MAC arrays, local SRAM buffers, and optimized control logic to maximize data reuse while minimizing energy-intensive memory traffic [6], [7]. Architectural strategies such as weight-stationary and row-stationary dataflows have been shown to reduce external memory bandwidth requirements, thereby achieving high levels of computational efficiency in low-power devices [8], [9]. Benchmarking studies confirm that NPUs achieve inference efficiency exceeding 3 TOPS/W in certain microcontroller-class deployments, surpassing both scalar cores and SIMD-based designs [6].

Despite these advances, much of the existing literature emphasizes either high-level algorithmic compression techniques or architectural descriptions of NPUs, without providing quantitative per-inference energy measurements under realistic embedded workloads. This omission is critical, as energy per inference directly determines battery life and deployment feasibility in real-world IoT and wearable applications. Moreover, few studies address cases where NPUs not only enhance efficiency but also enable execution of models otherwise infeasible on CPU-only platforms due to unsupported operations.

This work seeks to bridge this gap by combining (i) precise measurement methodology tailored for MCUs, (ii) experimental benchmarking of representative models across CPU and NPU execution paths, and (iii) detailed analysis of NPU utilization and energy efficiency. In doing so, it contributes empirical evidence supporting the role of NPUs as the cornerstone of energy-aware embedded ML deployments.

## III. MEASUREMENT METHODOLOGY

Optimization of machine learning models for microcontrollers is meaningful only when its impact can be quantified through precise and reproducible metrics. Unlike high-performance computing, where throughput metrics such as FLOPS or images per second dominate, embedded inference requires evaluation across a broader set of application-centric parameters, including inference latency, per-inference energy consumption, memory footprint, and accuracy preservation [10]. This section describes the measurement methodologies employed in this study, emphasizing practices suitable for resource-constrained systems.

### A. Inference Latency and Real-Time Performance

Inference latency, defined as the time between input arrival and output availability, is critical in applications such as gesture recognition and keyword spotting. Real-time feasibility typically requires latencies of only a few to several hundred milliseconds. Latency was measured using three methods:

On-chip timers, such as SysTick counters or hardware cycle counters.

External timing via GPIO toggling, where inference start and end signals were asserted on MCU pins and captured by an oscilloscope or logic analyzer.

Code instrumentation, using lightweight RTOS hooks or serial debug outputs.

Among these, GPIO-based timing was preferred due to its low overhead and high precision. Both average-case latency and worst-case latency were considered. The former guides model and system tuning, while the latter determines feasibility in safety-critical systems. Robustness was further validated across variations in supply voltage and environment temperature, ensuring consistent performance under real-world deployment conditions.

*B. Energy Consumption per Inference*

For battery-powered edge devices, energy per inference is the primary efficiency metric. It quantifies the electrical energy required to execute one inference, expressed in microjoules (μJ) or microwatt-hours (μWh). Unlike instantaneous power, which provides only a static snapshot, per-inference energy integrates both power consumption and execution time, yielding an application-relevant efficiency measure.

Energy was measured with an external high-precision instrument, specifically a Keysight 34465A Digital Multimeter (DMM), synchronized via GPIO triggers.

In our setup, each inference measurement was initiated by a software-generated GPIO pulse aligned with the inference window. The DMM captured current draw at high resolution, and energy was calculated by integrating power over the duration of the pulse. To isolate inference energy, idle-state power was separately measured and subtracted, yielding per-inference net energy. This methodology eliminates the confounding effects of background clocks and peripherals, providing a rigorous, reproducible basis for comparing CPU and NPU executions.

*C. Model Footprint and Memory Usage*

Memory footprint was characterized along two dimensions:
- Flash usage, including model weights and compiled binaries.
- RAM usage, including activations, scratch buffers, and stack allocations.

Flash usage was obtained directly from linker map files, while RAM usage was profiled using both static and dynamic methods. Static analysis was performed using compiler-generated. bbs outputs.

For frameworks such as TensorFlow Lite for Microcontrollers, RAM (dynamic) consumption is dominated by the tensor arena, a preallocated buffer that stores intermediate tensors and scratch memory. By summing linker reports with tensor arena allocations, we obtained reliable upper bounds on RAM usage prior to deployment. Accurate memory profiling is essential, as exceeding available SRAM can lead to silent inference failures or unstable system behavior.

*D. Summary*

The combination of latency, per-inference net energy, and memory footprint provides a holistic framework for assessing ML model optimization on microcontrollers. Among these, per-inference net energy is emphasized as the most representative metric for embedded AI deployments, as it captures the cumulative efficiency of both hardware and software optimization strategies. This methodology establishes the foundation upon which our experimental results and architectural evaluations in Sections IV and V are based.

## IV. Discrete Accelerators (NPUs)

Neural Processing Units (NPUs) represent specialized hardware accelerators designed to execute neural network inference with superior energy efficiency and throughput compared to general-purpose microcontroller cores. Unlike scalar pipelines or SIMD units, NPUs integrate dedicated multiply–accumulate (MAC) arrays, tightly coupled on-chip SRAM buffers, and optimized dataflow controllers, thereby minimizing memory traffic and maximizing performance-per-watt.

*A. Architectural Principles*

NPUs are typically constructed as arrays of Multiply-Accumulate (MAC) units organized into tiled clusters to support highly parallel execution of convolutional and fully connected layers. By collocating compute arrays with low-latency SRAM, NPUs significantly reduce off-chip memory accesses, which are otherwise a dominant source of energy consumption in embedded inference workloads. This architectural specialization enables NPUs to deliver inference performance in the order of 1 TOPS/W, far surpassing general-purpose CPU or SIMD implementations.

*B. Dataflow Strategies*

A central design element in NPUs is the orchestration of data movement. Different dataflow strategies optimize reuse of weights, activations, or partial sums.

Weight-stationary: Kernel weights are kept fixed in local SRAM, while activations are streamed across the compute array. This maximizes weight reuse and minimizes memory bandwidth

Output-stationary: Partial sums are retained in registers until an output activation is finalized, reducing memory transactions for intermediate results.

Row-stationary: Combines the above approaches by reusing both weights and partial sums simultaneously, achieving a balanced efficiency profile across diverse convolutional layers.

The choice of dataflow directly influences latency and energy consumption. For instance, weight-stationary designs, such as those pioneered in the Eyeriss accelerator, demonstrated significant improvements in CNN efficiency by reducing external memory traffic

*C. Sparsity Exploitation*

Modern NPUs frequently incorporate zero-skipping mechanisms that bypass computations involving zero-valued weights or activations, commonly produced by pruning or ReLU activations. Hardware-level sparsity support eliminates redundant MAC operations, reducing both latency and energy consumption with minimal area overhead. While zero-skipping requires additional scheduling logic, it is increasingly viewed as a standard feature in embedded NPUs due to its contribution to overall efficiency.

*D. Software Ecosystems*

Effective utilization of NPUs depends on compiler and runtime support. Toolchains translate high-level neural network graphs into optimized NPU instruction streams, performing tasks such as:
- Quantization mapping (e.g., INT8, INT4 precision support).
- Computation tiling to fit workloads into local SRAM.
- Scheduling operations to maximize MAC utilization while minimizing data transfers.

Frameworks such as TensorFlow Lite Micro provide NPU-aware backends that leverage these toolchains. Prior studies confirm that suboptimal graph-to-hardware mapping can diminish NPU efficiency [7] despite advanced hardware design. Our experiments also confirm that optimized conversion pipelines are essential for realizing full NPU benefits.

## V. EXPERIMENTAL RESULTS

To quantify the impact of Neural Processing Units (NPUs) on embedded machine learning performance, we conducted a set of controlled experiments on the Alif Semiconductor Ensemble E7 development board, which integrates an ARM Cortex-M55 MCU with the Ethos-U55 NPU. Six representative ML models of varying complexity were evaluated: MiniResNet, MobileNetV2, FD-MobileNet, MNIST, TinyYolo, and SSD-MobileNet. Each model was executed under two configurations: CPU-only inference (where CPU means the MCU core without the use of the NPU accelerator) and NPU-assisted inference. Measurements followed the methodology described in Section III, using GPIO-triggered Digital Multimeter (DMM) sampling synchronized with inference execution, with idle-state power subtracted to compute per-inference net energy.

*A. Sparsity Exploitation*

Table I summarizes inference latency, per-inference net energy, and corresponding improvement factors across CPU and NPU executions.

TABLE I
LATENCY AND ENERGY COMPARISON (CPU VS NPU ACROSS MODELS)

| Model | Latency (ms) CPU | Latency (ms) NPU | Latency Improve Factor | Net Energy (µWh) CPU | Net Energy (µWh) NPU | Energy Efficiency Factor |
|---|---|---|---|---|---|---|
| MiniResNet | 47.1 | 5.08 | 9.27x | 0.736 | 0.058 | 12.7x |
| MobileNet V2 | 320 | 8 | 39.6x | 5.949 | 0.112 | 53.1x |
| FD-MobileNet | 73.9 | 10.68 | 6.91x | 1.239 | 0.139 | 8.9x |
| MNIST | 3.22 | 5.16 | 0.62x | 0.038 | 0.058 | 0.65x |
| TinyYolo | 10589 | 83.6 | 126.4x | 191.673 | 1.399 | 143.1x |
| SSD-MobileNet | (unsupported) | 10.65 | – | – | 0.139 | – |

Results show that NPU acceleration consistently reduced latency and energy consumption for medium- and large-scale models. TinyYolo achieved the most dramatic benefit, with latency reduced by over 125× and energy consumption by 143×. By contrast, MNIST exhibited a rare case of performance regression, with slightly higher latency and energy on the NPU, caused by underutilization overhead at only 5.1% NPU activity. Importantly, SSD-MobileNet, which failed on the CPU due to an unsupported operation (TILE opcode), executed efficiently on the NPU, underscoring the functional advantages of dedicated accelerators.

*B. Memory Footprint and Model Deployability*

Table II reports memory requirements, including activation buffer size (RAM) and total model size (Flash). These values are critical for deployment feasibility in microcontrollers with limited resources.

TABLE II
MEMORY FOOTPRINT AND ACTIVATION BUFFER USAGE

| Model | Activation Buffer (Bytes) | Model Size (Bytes) | Notes |
|---|---|---|---|
| MiniResNet | 110,480 | 141,664 | Small CNN, efficient on NPU |
| MobileNet V2 | 1,005,040 | 712,480 | Large model, near-maximum utilization |
| FD-MobileNet | 1,005,952 | 200,256 | Mid-complexity, efficient scaling |
| MNIST | 17,008 | 21,712 | Very small, underutilizes NPU |
| TinyYolo | 1,879,504 | 9,700,352 | High complexity, best NPU gains |
| SSD-MobileNet | 1,321,719 | 1,007,712 | CPU-unsupported, feasible with NPU |

The data highlight two important points. First, large activation buffers are required by advanced models such as MobileNetV2 and TinyYolo, underscoring the importance of sufficient on-chip SRAM or fast external memory interfaces. Second, even large and unsupported models (e.g., SSD-MobileNet at ~1 MB) can be executed on MCU-class devices when supported by NPUs, expanding the feasible range of embedded AI workloads.

*C. NPU Utilization and Internal Activity Metrics*

Table III presents detailed NPU utilization metrics, including active/idle cycles and memory read/write activity. These values provide insight into how efficiently the NPU executes different workloads.

TABLE III
NPU UTILIZATION AND INTERNAL ACTIVITY METRICS

| Model | NPU Utilization (%) | NPU Active Cycles | NPU Idle Cycles | Memory Reads | Memory Writes |
|---|---|---|---|---|---|
| MiniResNet | 49.5 | 365962 | 373319 | 87814 | 43735 |
| MobileNet V2 | 99.9 | 2102495 | 1436 | 815559 | 556383 |
| FD-MobileNet | 49.7 | 734143 | 743979 | 326138 | 215099 |
| MNIST | 5.1 | 38055 | 701185 | 8291 | 5210 |
| TinyYolo | 100 | 32250950 | 1181 | 2997071 | 1472289 |
| SSD-MobileNet | 49.8 | 736007 | 742070 | - | - |

The data reveal strong correlations between utilization and efficiency. High-utilization models (MobileNetV2, TinyYolo) achieved the largest energy and latency reductions, validating the architectural specialization of the Ethos-U55. Conversely, MNIST's very low utilization explains its inefficiency when offloaded to the NPU.

*D. Key Observations*

Performance scales with complexity: The greater the computational intensity of the model, the higher the NPU utilization and the greater the efficiency gains.

NPUs enable new workloads: SSD-MobileNet could not run on the CPU at all, yet executed efficiently on the NPU.

Overhead exists for tiny models: Lightweight networks like MNIST may not benefit from NPU acceleration, as the fixed cost of NPU orchestration outweighs computational savings.

## VI. DISCUSSION

The experimental evaluation confirms that Neural Processing Units (NPUs) play a decisive role in enabling efficient machine learning inference on microcontrollers. Results demonstrate that the efficiency of NPU acceleration scales strongly with model complexity. While small networks such as MNIST showed limited or even negative benefits due to low utilization overhead, more complex architectures such as MobileNetV2, TinyYolo, and SSD-MobileNet achieved utilization levels close to 100%, yielding dramatic reductions in both latency and energy consumption. In the case of TinyYolo, for example, latency decreased by more than 125 times and per-inference net energy by 143 times relative to CPU-only execution.

Beyond efficiency, NPUs also provide functional enablement. The SSD-MobileNet model, which could not be executed on the CPU due to unsupported instructions, ran efficiently on the Ethos-U55 NPU, with latency under 11 ms and energy consumption of only 0.139 µWh per inference. This highlights NPUs' ability not only to accelerate supported workloads but also to extend the range of models that are deployable on MCU-class devices.

The experiments further validate per-inference net energy as the most meaningful evaluation metric for embedded ML. Unlike instantaneous power, which during our measurements showed modest differences between CPU and NPU runs, net energy captured the combined effect of latency and utilization. In practice, slightly higher instantaneous power was more than offset by drastically shorter execution times, producing large reductions in total energy. This confirms that energy-aware deployment decisions must consider execution time alongside raw power consumption.

At the architectural level, the results emphasize the importance of efficient memory hierarchies and dataflow scheduling. Models requiring large activation buffers, such as MobileNetV2 and TinyYolo, stressed SRAM capacity and bandwidth; nevertheless, the NPU sustained efficiency by minimizing redundant memory transfers, as reflected in its high active-to-idle cycle ratios. Conversely, models with lower arithmetic intensity achieved only partial utilization, limiting efficiency gains. These observations reinforce that data movement—not arithmetic throughput—remains the primary bottleneck in embedded inference.

Finally, the findings highlight the necessity of hardware–software co-design. The ability of the NPU toolchain to successfully remap unsupported CPU operations in SSD-MobileNet exemplifies the critical role of compilers and runtime frameworks. Without optimized graph-to-hardware mapping, even advanced accelerators risk underutilization. From a system design perspective, this suggests a tiered acceleration strategy. Lightweight models may execute efficiently on CPUs with caching and SIMD support, while medium- to large-scale models require NPU offload to achieve real-time, energy-efficient inference.

In summary, the results demonstrate that NPUs substantially expand the boundaries of feasible embedded AI by reducing latency, minimizing energy consumption, and enabling execution of advanced workloads. Efficiency improvements scale with utilization, memory efficiency is as critical as computational throughput, and compiler support is essential to unlock the full potential of hardware acceleration. These insights establish NPUs as a cornerstone of next-generation MCU-based AI systems.

## VII. CONCLUSIONS

This work has presented a detailed evaluation of hardware acceleration for machine learning inference on microcontrollers, with emphasis on the role of Neural Processing Units (NPUs). Using a rigorous per-inference net energy methodology, we compared CPU-only and NPU-assisted executions across six representative models. The results confirm that hardware acceleration is not merely an optimization but a foundational requirement for energy-efficient embedded AI.

Three key insights emerged. First, NPUs provide orders-of-magnitude improvements in both latency and energy for medium- and high-complexity models. For TinyYolo, the NPU reduced per-inference net energy by 143× and latency by 126×, while also enabling execution of SSD-MobileNet, which was infeasible on the CPU. Second, efficiency scales with utilization: high utilization correlated with significant gains, while small models such as MNIST showed minimal benefit due to orchestration overheads. Third, memory hierarchies and software co-design are equally critical. Activation buffer size and dataflow scheduling strongly influenced performance, and NPU efficiency was contingent on optimized compiler support for graph-to-hardware mapping.

These findings suggest a tiered acceleration strategy: lightweight workloads may execute efficiently with caching and SIMD support, while advanced architectures require NPUs to achieve real-time, energy-constrained inference. More broadly, the results demonstrate that per-inference net energy is the most reliable metric for evaluating embedded ML efficiency, as it integrates both power draw and execution time into a single measure relevant for battery-powered deployments.

Looking forward, the integration of NPUs with capable MCU cores, robust memory hierarchies, and mature software toolchains will redefine the performance envelope of edge devices. Hardware acceleration has thus transitioned from an optional enhancement to a cornerstone of modern microcontroller design, enabling a new generation of intelligent, energy-aware, real-time embedded systems.

ACKNOWLEDGMENT

………